# A Learning-Based 3D EIT Image Reconstruction Method

Zhaoguang Yi[1], Zhou Chen[1], and Yunjie Yang[1]

[1]Institute for Digital Communications, School of Engineering, the University of Edinburgh, UK

Correspondence: Zhaoguang Yi, e-mail: yizhaoguang@outlook.com

**Abstract**– Deep learning has been widely employed to solve the Electrical Impedance Tomography (EIT) image reconstruction problem. Most existing physical model-based and learning-based approaches focus on 2D EIT image reconstruction. However, when they are directly extended to the 3D domain, the reconstruction performance in terms of image quality and noise robustness is hardly guaranteed mainly due to the significant increase in dimensionality. This paper presents a learning-based approach for 3D EIT image reconstruction, which is named Transposed convolution with Neurons Network (TN-Net). Simulation and experimental results show the superior performance and generalization ability of TN-Net compared with prevailing 3D EIT image reconstruction algorithms.

## 1. Introduction

Electrical Impedance Tomography (EIT) has subscribed to the belief that it is a non-intrusive and non-radiative imaginging modality in clinical and industrial sectors. In EIT, the conductivity distribution within the region of interest can be reconstructed using boundary voltage measurements induced by current injections. The EIT-image-reconstruction problem is a long-standing challenge in the field due to its ill-posed and ill-conditioned nature (Yang, 2018). Recently, researchers have made noteworthy progress on both physical model-based, and learning-based 2D EIT image reconstruction algorithms (Yang *et al* 2014, Chen *et al* 2020). However, the EIT image reconstruction problem especially in 3D remains to be thoroughly investigated. Directly transferring from 2D to 3D based on the developed approaches usually requires extensive computational resources. Additionally, further improvements with respect to image quality and noise reduction performance are required.

This paper proposes a Transposed convolution with Neurons Network (TN-Net) for the EIT-image-reconstruction in 3D. The proposed network structure is inspired by the 3D Generative Adversarial Network (3D-GAN) (Wu *et al* 2016). Before the transposed convolution layers, a decoder is introduced, which can decode the normalized EIT measurement data to a more representative version in probability domain. The decoder is constructed by a stack of fully connected layers, which add a portion of buffering to the transposed convolutional layer compared to passing the measurement data directly, making the network perform better. After mapping the measurement domain to a low-dimensional probability domain based on the decoder, the subsequent 3D transposed convolution layers complete the interpretation from the low-dimensional probabilistic space to the high dimensional space representing 3D reconstructed results (Wu *et al* 2016).

Three main contributions of this work are:
1. A network architecture TN-Net for 3D EIT image reconstruction is proposed. The decoder conducts structural learning which tackles the nonlinearity of the inverse problem. The following transposed convolutional layers further enhances the image quality of 3D reconstructions. Using the same computing resources, the TN-Net enables higher-quality 3D EIT image reconstruction compared with non-structural learning..
2. A 3D EIT dataset is constructed, which comprises 21,135 randomly generated multi-object, multi-conductivity level phantoms.
3. The TN-Net is verified using simulation and experimental data, demonstrating its superior performance in terms of computational efficiency, noise robustness, and generalization ability over the exisiting reconstruction algorithms.

## 2. Methodology

### 2.1 Electrical Impedance Tomography

The forward problem of EIT is to reconstruct the conductivity distribution from induced boundary voltage measurements. The commonly used Complete Electrode Model (CEM) (Holder, 2004) for EIT is described by:

$$\nabla \cdot (\sigma(x,y)\nabla \mu(x,y)) = 0, (x,y) \in \Omega \tag{1}$$

$$u + Z_l \sigma \frac{\partial u}{\partial n} = U_l, l = [1, N] \tag{2}$$

$$\int_{e_l} \sigma \frac{\partial u}{\partial n} dS = I_l, l = [1, N] \tag{3}$$

$$\sigma \frac{\partial u}{\partial n} = 0 \tag{4}$$

where $N$ is the number of electrodes, $\sigma$ is the conductivity distribution, $\partial \Omega$ is the the boundary, $e_l$ is the $l$th electrode, and $u$ is the potential on the boundary, which is excited by the injected current $I$. The complete electrode model can be solely solved by using the conservation of charges and the arbitrary choice of reference potentials, i.e. $\sum_{l=1}^{N} I_l = 0$, $\sum_{l=1}^{N} u_l = 0$.

The inverse problem of EIT is to estimate an unknown image $\sigma$ from given voltage measurements $V$. In this work, without requiring any a priori knowledge, the nonlinear relationship between the perturbation of conductivity distribution $\Delta \sigma$ and the induced voltage measurement changes $\Delta V$ is learned via a deep network.

### 2.2 Network Structure

TN-Net is developed to reconstruct 3D EIT images from boundary voltage measurements. The network structure is shown in Figure 1. TN-Net consists of two parts: decoding and reconstructing parts.

Our trials show that the expressive power of transposed convolutional layers was insufficient to solve the inverse problem of 3D EIT. We propose to add a decoding part using three fully connected layers, with feature sizes of 256,512, and 1024, respectively. Each fully connected layer is followed by a dropout layers. During training, before feeding the input data into the network, we introduce an independent noise addition module to add noise of 35dB SNR to the input data, which drives the model to learn eternal patterns from all kinds of noise-contaminated data, thus increasing its generalizability.

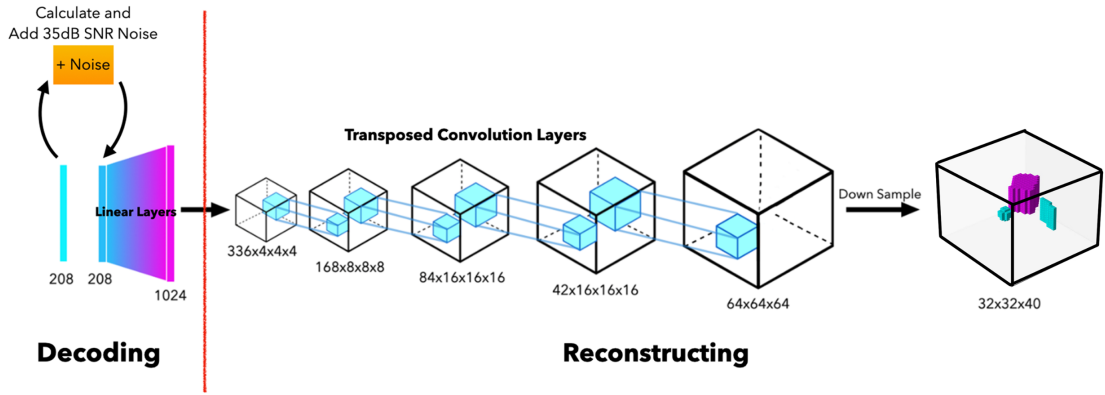

**Figure 1.** The architecture of TN-Net.

The reconstructing part is inspired by the 3D-GAN (Wu *et al* 2016), which contains 4 transposed convolution layers with convolution kernels of (4,4,4) and stride of (2,2,2). Each transposed convolution layer is paired with a batch normalization layer and a Leaky ReLU layer. Finally, the resulting reconstructed conductivity distribution with a high-resolution of 64*64*64 is downsampled to 32*32*40. The source code of TN-Net is available from: https://github.com/Josh-Yi/TN-Net.

## 3. Experimental Setup

### 3.1 3D EIT Dataset

The 3D EIT dataset is generated using COMSOL Multiphysics by solving the forward problem of 3D EIT using the Finite Element Method (FEM). As shown in Fig. 2, the COMOSOL model was built based on the experimental equipment, consisting of two 16-electrode layers. We adopt the adjacent measurement strategy (Yang et al 2017), and the completed non-redundant measurement cycle consists of 208 voltage measurements. The targeted 3D image to be reconstructed is a 32*32*40 conductivity distribution. The measurement-image pairs are calibrated and normalized to eliminate common errors caused by systematic defects. Four types of phantoms (see Fig. 3) are generated with randomly selected shape, size, rotation angle, conductivity, and number of objects for the dataset according to the generation rules listed in Table 1.

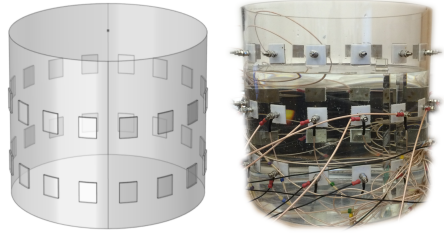

**Figure 2.** The 3D model of the 32 electrode EIT sensor and the real-world sensor.

**Table 1.** *3D EIT* Dataset Generation Rules.

| Total | 2 Objects - | 2 Objects ± | 3 Objects - | 3 Objects ± |
|---|---|---|---|---|
| **21,135** | 4,352 | 4,520 | 7,201 | 5,062 |

*- indicates that the normalized conductivity change is negative, while ± means the phantoms have both negative and positive conductivity changes.*

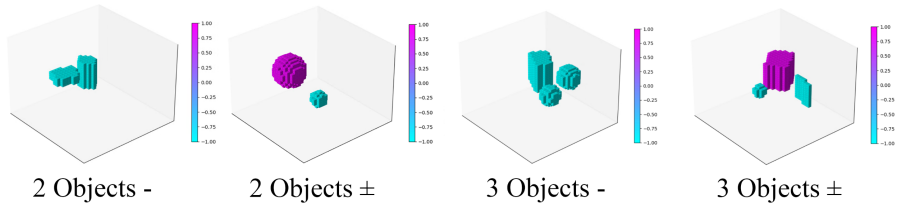

2 Objects -    2 Objects ±    3 Objects -    3 Objects ±

**Figure 3.** Four representative examples in the 3D EIT dataset.

### 3.2 Network Training

As mention in Section 2.2, an additive noise of 35dB SNR is added to the EIT measurement in the training process in a way that the model continuously sees different noise-contaminated samples during each epoch. Dynamically adding the noise can increase the model's robustness and prevent overfitting. The optimizer is the Adam (Loshchilov, 2019) with a learning rate of 0.002, a weight decay of 0.01, and betas of (0.9, 0.999). The model is trained 300 epochs with a batch size of 442. The model at 260-th epoch achieved the minimal validation loss and was selected as the final model.

## 4. Results

We compare the proposed TN-Net with two 3D EIT reconstruction algorithms, i.e., one-step Gauss-Newton solver with Laplace Regularization (One-Step) (Yang et al 2014) and the original generator of 3D-GAN (Wu et al 2016).

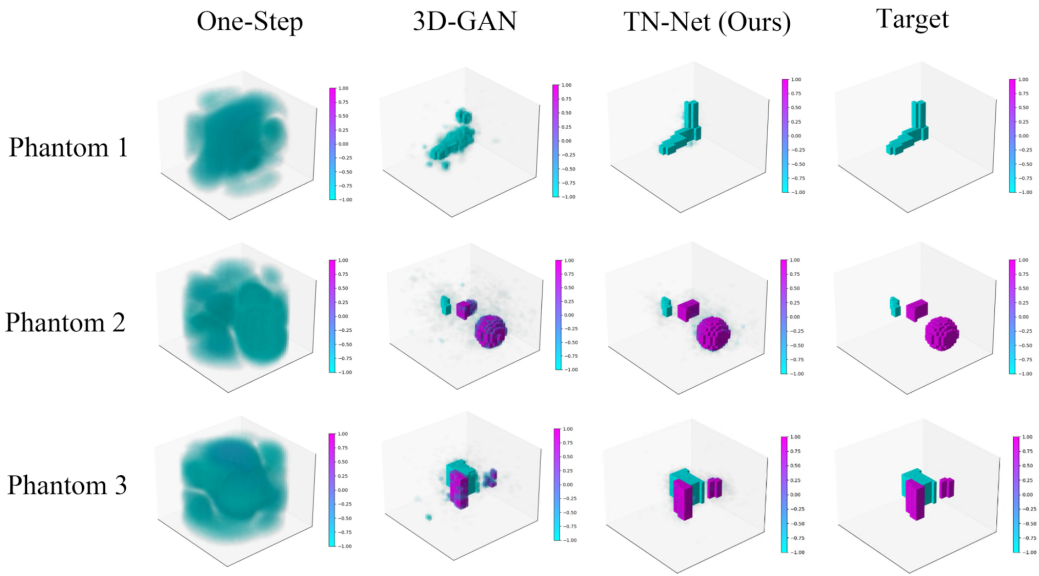

**Figure 4.** 3D EIT reconstruction results of unseen simulation data.

The test data are added with 30 dB noise worse than the training process. Some representative results from the testing set are shown in Fig. 4. Obviously, our TN-Net stands out compared with the other two methods. It has the best noise resistance performance by reconstruting the best positions and shapes of all inclusions. Furthermore, TN-Net is capable of better distinguishing phantoms with both positive and negative conductivity changes, outperforming the other methods.

Table 3 reports quantitative comparisons of all methods over the testing data based on the statistic average of Root Means Square Error (RMSE), 3D Structural Similarity Index (SSIM), Peak Signal to Noise Ratio (PSNR), and inference time. Apart from the best RMSE, SSIM, and PSNR, the TN-Net has higher imaging speed than the traditional one-step method and is slightly slower than 3D-GAN due to the deeper architecture of TN-Net with an additional decoder. Nevertheless, the execution time of TN-Net (below 40 ms) is sufficiently good for implementing real-time 3D EIT imaging.

**Table 2.** Result evaluation on the whole test dataset.

|  | RMSE | SSIM | PSNR | Inference Time |
|---|---|---|---|---|
| Ours | **7.616e-5** | **0.9657** | **30.709dB** | 0.0311s |
| One-Step | 0.1267 | 0.6193 | 21.847dB | 0.0478s |
| 3D-GAN | 9.176e-5 | 0.9485 | 29.082dB | **0.0306s** |

The average inference time is tested on an Intel i9-12900k. The best results are highlighted in bold.

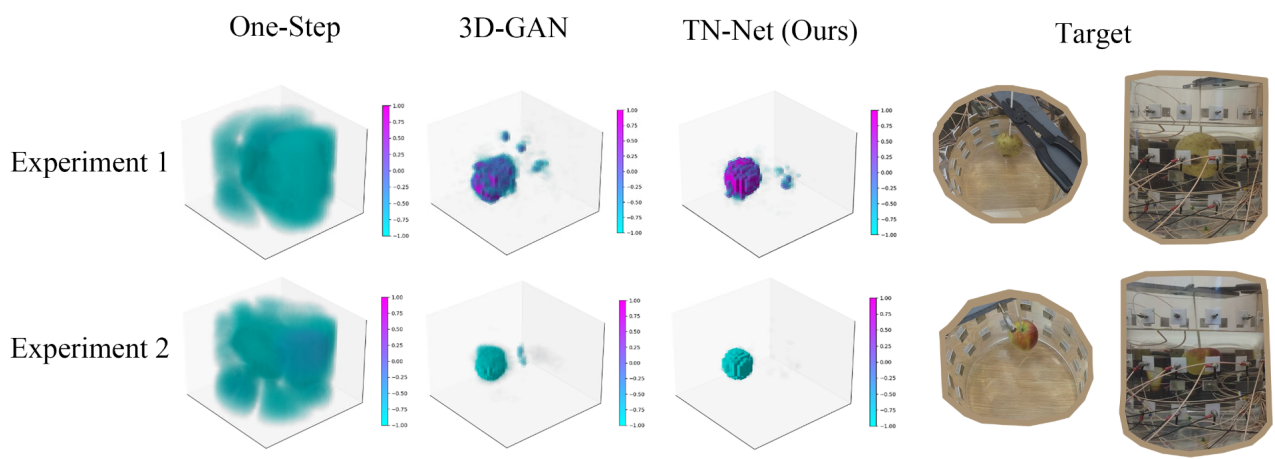

**Figure 5.** 3D EIT reconstruction results using experimental data.

We also validate the proposed method using real-world data (see Fig. 5). Two phantoms are reconstructed with current injection of 25 kHz and a 3D EIT sensor with 16*2 electrodes. The potato in Experiment 1 is a tuber with higher ionic and water content than Experiment 2 with an apple, which is successfully recognized by 3D-GAN and TN-Net. The experimental results further verified the generalization ability and superior performance of the TN-Net over the other methods.

## 5. Conclusion

This paper proposes a novel learning-based 3D-EIT reconstruction method named TN-Net. TN-Net has demonstrated its robustness and superior performance on both simulation and real-world 3D EIT data. TN-Net is around 40% faster and 60% more accurate in terms of SSIM than the comparing algorithms. Future work will investigate the application of the proposed TN-Net in 3D tissue imaging.